 \documentstyle[multicol,aps,prl,psfig]{revtex}
\newcommand{\be}{\begin{equation}}
\newcommand{\ee}{\end{equation}}
\newcommand{\ba}{\begin{eqnarray}}
\newcommand{\ea}{\end{eqnarray}}

\newcommand{\br}{{\bf r}}
\newcommand{\bk}{{\bf k}}

\newcommand{\eps}{\epsilon}

\begin{document}
\title{Mesoscopic Theory of Granular Fluids}
\author{
T.P.C.~van Noije and M.H.~Ernst\\
{\it Instituut voor Theoretische Fysica, Universiteit Utrecht, Postbus 80006,
3508 TA Utrecht, The Netherlands}
\\
\vspace*{5pt}
R.~Brito and J.A.G.~Orza\\
{\it Facultad de Ciencias F\'{\i}sicas, Universidad Complutense,
28040 Madrid, Spain}}
\date{to be published in Physical Review Letters}
\maketitle
\begin{abstract}
Using fluctuating hydrodynamics we describe the slow build-up
of long range spatial correlations
in a freely evolving fluid of inelastic hard spheres.
In the incompressible limit, the behavior of spatial velocity correlations
(including $r^{-d}$-behavior) is governed by vorticity fluctuations
only and agrees well with two-dimensional simulations up to 50 to
100 collisions per particle.
The incompressibility assumption breaks down beyond a distance that
diverges in the elastic limit.
\end{abstract}
\pacs{05.20.Dd, 05.40.+j, 81.05.Rm}
\begin{multicols}{2}
\narrowtext
In the characterization of granular matter as un unusual solid,
fluid or gas by Jaeger et al.\ \cite{jaeger}, this letter addresses
the {\em granular gas} regime, controlled by inelasticity,
clustering \cite{goldhirsch} and collapse \cite{mcnamara}.
Clustering is a long wavelength, low frequency (hydrodynamic)
phenomenon and inelastic collapse a short wavelength, high
frequency (kinetic) phenomenon.
In the granular gas regime, also called {\em rapid granular flows},
the dynamics is dominated by inelastic collisions.
Here the methods of nonequilibrium statistical mechanics, molecular
dynamics, kinetic theory and hydrodynamics are most suitable for
describing the observed average macroscopic behavior
\cite{goldhirsch,mcnamara,deltour,esipov,brey,boston} and the
fluctuations around it.

The lack of energy conservation makes the granular gas, whether
driven or freely evolving, behave very differently from molecular
fluids.
The essential physical processes and detailed dynamics are
described in \cite{goldhirsch,mcnamara} and references therein:
the similarities and differences with molecular fluids; lack of
separation of microscales and macroscales, not only because the
grains themselves are macroscopic, but also because of the
existence
of intermediate intrinsic scales which are controlled by the
inelasticity and are only well separated when the system is nearly
elastic.
A simple model which incorporates the inelasticity of the granular
collisions, consists of inelastic hard spheres (IHS), taken here of
unit mass and diameter, with momentum
conserving dynamics.
The energy loss in a collision is proportional to the {\em
inelasticity
parameter} $\epsilon=1-\alpha^2$ where
$\alpha$ is the coefficient of normal restitution.

For an understanding of what follows we recall two important
properties of the undriven granular gas: (i) the existence of a
{\em homogeneous cooling state} (HCS) and (ii) its instability
against spatial fluctuations.
The hydrodynamic equations for an IHS-fluid, started in a uniform
equilibrium state with temperature $T_0$, admit an HCS-solution (see
e.g.\ \cite{goldhirsch,mcnamara,boston}) with a homogeneous
temperature $T(t)$, described by $\partial_t T=-2\gamma_0 \omega T$.
Here the collision frequency is $\omega(T)\sim \sqrt{T}/l_0$ with a
mean free path $l_0$, given by the Enskog theory \cite{chapman} for
a {\em dense} system of hard disks or spheres ($d=2,3$) and
$\gamma_0=\epsilon/2 d$.
Then $T(t)=T_0/[1+\gamma_0 \omega(T_0) t]^2 =T_0
\exp(-2 \gamma_0 \tau)$, where $\tau$ is the average number of collisions
suffered per particle within a time $t$.
It is found by integrating $d\tau =\omega (T(t)) d t$.
Moreover, this HCS-solution is {\em linearly unstable} once the
linear extent $L$ of the system exceeds some dynamic correlation
length, which increases with decreasing $\eps$, and is proportional
to $l_0$ \cite{jaeger,goldhirsch,mcnamara,deltour,esipov,brey,boston}.

The dynamics of {\em fluctuations}, say, in
density, $\delta n (\br,t)$, and flow field, ${\bf
u}(\br,t)$, have hardly been studied 
\cite{goldhirsch,deltour,boston}, in sharp contrast to the
large number of publications about the
average behavior. 
We note that fluctuations are absent in hydrodynamic as well as in
Boltzmann-Enskog-type kinetic equations which are based on
molecular chaos (mean field assumption).
The objects of interest in this letter are the spatial velocity 
and density correlations
\ba
G_{\alpha\beta}(\br,t)&=&\frac{1}{V} \int d{\bf r}^\prime \langle 
u_\alpha({\bf r}+\br^\prime,t) u_\beta({\bf r}^\prime,t)\rangle\nonumber\\
G_{nn}(\br,t)&=&\frac{1}{V}\int d{\bf r}^\prime
\langle \delta n({\bf r}+\br^\prime,t) \delta n({\bf r}^\prime,t)\rangle,
\ea
with
$V=L^d$, and
the structure factors $S_{\alpha \beta}(\bk,t)$ and
$S_{nn}(\bk,t)$, which are the corresponding Fourier transforms.
Goldhirsch et al.\ \cite{goldhirsch} initiated molecular dynamics
studies of $S_{nn}(\bk,t)$ and $S_{pp}(\bk,t)=\sum_\alpha
S_{\alpha\alpha}(\bk,t)$, and related in a qualitative way the
structure at small $k$ to the most unstable vorticity modes, and
presented a nonlinear analysis to explain the enslaving of density
fluctuations by the vorticity field \cite{footnote}.
A more quantitative description of the structure factors
$S_{nn}(\bk,t)$ \cite{deltour} and $S_{pp}(\bk,t)$ \cite{boston}
has been recently proposed, based on the dynamics
of macroscopic unstable modes
(Cahn-Hilliard theory of spinodal decomposition \cite{langer}).
However, numerical evidence from molecular dynamics for the
quantitative validity of this theory is still lacking. 

The main goal of the present paper is to calculate the velocity
correlation function $G_{\alpha\beta}(\br,t)$ in unforced granular
flows and to
show that fluctuating hydrodynamics \cite{landau}
gives a quantitative description of the spatial correlations 
over a large intermediate time interval,
controlled by linearized hydrodynamics.
Because the flows in freely evolving IHS systems are approximately {\em
incompressible} ($\mbox{\boldmath $\nabla$}\cdot {\bf u}=0$), the vorticity
field, $\mbox{\boldmath $\nabla$} \times {\bf u}$, in the nonlinear
Navier-Stokes equations is practically
decoupled from the other hydrodynamic fields in the system.
This implies that the density $n$ and the temperature
$T(t)=T_0\exp(-2 \gamma_0 \tau)$ can be considered homogeneous, and an
approximate theory based on vorticity fluctuations alone is
justified.

Thus, we describe the Fourier modes of the vorticity field or
transverse flow field ${\bf u}_\perp(\bk,t)$ by the mesoscopic
Langevin equation \cite{landau} (valid for $k l_0 \lesssim 1$)
\be
\partial_t {\bf u}_\perp(\bk,t)+\nu(T(t)) k^2 {\bf u}_\perp(\bk,t)=
\hat{\bf F}(\bk,t),
\label{eq:1}
\ee
where ${\bf u}_\perp$ and $\hat{\bf F}$ are orthogonal to $\bk$, and
where
$\nu(T)\sim l_0 \sqrt{T}$ is the kinematic viscosity of the IHS-fluid
in the HCS.
The random noise $\hat{\bf F}$ is assumed to be white and
Gaussian with a correlation
\be
\langle \hat{F}_\alpha(\bk,t)\hat{F}_\beta(-\bk,t^\prime)\rangle/V=
B_{\alpha\beta}(T(t)) k^2 
\delta(t-t^\prime),
\label{eq:2}
\ee
and a noise strength $B_{\alpha\beta}(T)=2 \delta_{\alpha\beta} T
\nu(T) /n$ \cite{landau}.
With the additional assumption (see \cite{goldhirsch,mcnamara}) that the 
IHS-viscosity has the same functional form as for elastic hard
spheres or disks, the solution of the proposed Langevin theory
provides a detailed prediction for $G_{\alpha\beta}(\br,t)$ on
hydrodynamic space ($r\gtrsim l_0$) and time scales 
($\tau\gtrsim 1/\omega(T_0)$) {\em without any}
adjustable parameters.

We briefly indicate how this is done by calculating the structure
factor $S_\perp(\bk,t)=\langle|u_{\perp\alpha}(\bk,t)|^2\rangle/V$.
Here the subscript $\alpha$ is one of the $(d-1)$ equivalent transverse
components of ${\bf u}_\perp$.
We transform Eq.\ (\ref{eq:1}) into the standard Langevin equation
with {\em time independent} noise strength and coefficients. 
This is done by the change of variables $\omega(T(t))dt=d\tau$,
${\bf u}_\perp(\bk,t)=\sqrt{T(t)} {\bf w}(\bk,\tau)$ and
$\hat{\bf F}(\bk,t)=\omega(T(t))\sqrt{T(t)} \hat{\bf f}(\bk,\tau)$ and
yields $\partial_\tau {\bf w}(\bk,\tau)-z_\perp(k){\bf
w}(\bk,\tau) = \hat{\bf f}(\bk,\tau)$, with a growth rate
$z_\perp(k)=\gamma_0 (1-k^2 \xi^2)$ and a noise strength 
$b_{\alpha\beta}=
2 \delta_{\alpha\beta}\gamma_0 \xi^2 /n$. 
The dynamic correlation length $\xi\equiv \sqrt{\nu/\omega
\gamma_0}$ is time independent and of order $
l_0/\sqrt{\gamma_0}$.
With the help of the relation $\langle|w_\alpha(\bk,0)|^2\rangle=V/n$,
the structure factor is then found as
\be
S_\perp(k,t)=\frac{T(t)}{n}\left\{1+\frac{\exp[2\gamma_0
\tau(1-k^2\xi^2)]-1}{1-k^2\xi^2}\right\}
\ee
which is valid for $k l_0\lesssim 1$.
In the elastic limit ($\gamma_0\rightarrow 0$) the standard form of
fluctuating hydrodynamics and the fluctuation dissipation theorem are
recovered.
For $k\xi\lesssim 1$ this equation describes {\em new structure} of
the velocity correlations $G_{\alpha\beta}(\br,t)$ on length scales
of order $2 \pi \xi$.
At the end of the paper we return to the predicted structure on the
largest scales.

On the shortest scales ($r\rightarrow0$),
$G_{\alpha\beta}(\br,t)\rightarrow [T(t)/n]\delta_{\alpha\beta}
\delta(\br)$, caused by self-correlations of particles.
As our theory describes only structure on the scale $r\gtrsim l_0$, we
consider the equivalent function $G^+_{\alpha\beta}(\br,t)$ with
the self-correlations substracted, which is regular at the origin.
For the same reason the structure factor $S_{\alpha\beta}(\bk,t)$
has a plateau value $\delta_{\alpha\beta} T(t)/n$ for
$k\rightarrow \infty$, whereas $S^+_{\alpha\beta}(\bk,t)\rightarrow
0$ in the same limit.
The function $S^+_{\alpha\beta}(\bk,t)$ is an isotropic tensor
field, which can be decomposed into two independent scalar
functions of $k=|\bk|$, i.e.\
$S^+_{\alpha\beta}(\bk,t)=\hat{k}_\alpha\hat{k}_\beta
S^+_\parallel(k,t)+(\delta_{\alpha\beta}-\hat{k}_\alpha\hat{k}_\beta)
S^+_\perp(k,t)$, where a hat denotes a unit vector.
The {\em incompressibility assumption} implies then
$u_\parallel(\bk,t)=0$, and consequently $S^+_\parallel(k,t)=0$.
Similarly, $G^+_{\alpha\beta}(\br,t)=\hat{r}_\alpha\hat{r}_\beta
G^+_\parallel(r,t)+(\delta_{\alpha\beta}-\hat{r}_\alpha \hat{r}_\beta)
G^+_\perp(r,t)$.
Consider first the longitudinal spatial correlation
$G^+_\parallel(r,t)\equiv
[T(t)/n \xi^d] g_\parallel(r/\xi;2\gamma_0
\tau)$, which is given by
\be
g_\parallel(x,s)=\int \frac{d{\bf q}}{(2\pi)^d} e^{i{\bf q}\cdot{\bf
x}}\sin^2{\theta} \frac{\exp[s(1-q^2)]-1}{1-q^2},
\ee
where $\cos{\theta}=\hat{\bf q}\cdot\hat{\bf x}$.
In the {\em incompressible limit} the transverse correlation
function is given by $g_\perp(x,s)=g_\parallel(x,s)+[x/(d-1)]
\partial g_\parallel(x,s)/\partial x$ (see \cite{landau}, chapter
3).
These functions can be expressed as integrals over simple
functions.
As an example, we quote the result for $d=2$:
\be
g_\parallel(x,s)=\frac{1}{2\pi x^2}\int_0^s ds^\prime \exp(s^\prime)
[1-\exp(-x^2/4 s^\prime)].
\ee
The transverse function $g_\perp(x,s)$ has a negative minimum;
moreover $g_\parallel(x,s)$ is {\em positive} for all $x, s, d$;
there are {\em algebraic} tails $g_\parallel(x,s)\sim
-(d-1)g_\perp(x,s)\sim x^{-d}$ with a correction term of ${\cal
O}(\exp(-x^2/4s))$.
Similar algebraic tails occur in nonequilibrium stationary states
in driven diffusive systems \cite{grinstein}.
These functions have structure on hydrodynamic space and time scales
where both $x=r/\xi$ and $s=2\gamma_0\tau$ can be either large or
small with respect to unity.
At small inelasticity ($\gamma_0\rightarrow 0$) the dynamic
correlation length and mean free path $l_0$ are well separated.
Details will be published elsewhere \cite{brito}.

To verify the theory quantitatively we have
performed event-driven molecular dynamics simulations of smooth
inelastic hard disks, using
square periodic boundary conditions and $N=5\times 10^3$, $2\times
10^4$ and $5\times 10^4$ particles.
The inelasticity parameter $\epsilon=1-\alpha^2$ was varied between
0.02
and 0.8, and the area fraction from
$\phi=0.02$ to 0.4, far below
the solid transition ($\phi_{\rm solid}=0.665$).
Before considering the range of validity of our results,
we show in Fig.\ 1a how the relative vorticity fluctuations
${\bf w}(\bk,\tau)$ have grown \cite{goldhirsch,mcnamara,boston}.
The HCS and linear hydrodynamics start to break down.
The vorticity field becomes large and evolves into a `dense fluid
of closely packed vortex structures', which is still homogeneous
on scales large compared to $L_v$.
This happens the more rapidly, the larger the
inelasticity parameter $\epsilon$.

\begin{figure}
\[ \psfig{file=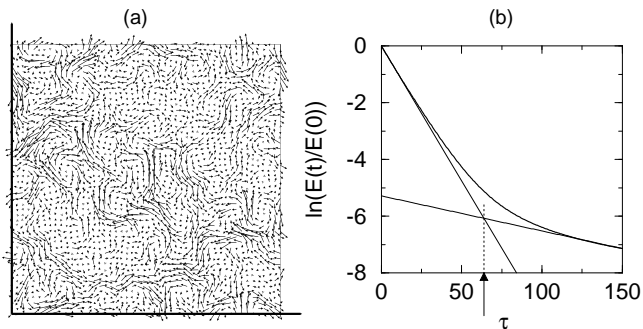} \]
\caption{
(a) Snapshot of the momentum density at $\tau=80$ in a system with
$N=5\times 10^4$, $L=313$ ($\phi=0.4$), $\alpha=0.9$,
coarse grained over cells of $6.3\times 6.3$,
representing a rather stable (still very similar at $\tau=160$)
 configuration of vortex
structures of typical diameter $L_v(t)\simeq 25$ ($l_0\simeq 0.34$)
with nearest
neighbors having opposite vorticity for reasons of stability.
(b) $\ln[E(t)/E(0)]$ versus $\tau$ from the same simulation showing
the linear (HCS) regime, where $E(t)=T(t)$,
with slope $-2\gamma_0$,
up to a crossover time $\tau_{\rm cr}\simeq 65$,
and the nonlinear cooling regime with a smaller slope,
where the cooling through inelastic collisions is
partially compensated by viscous heating.}
\end{figure}

Apart from the restrictions to hydrodynamic space and
time scales, there are two essential criteria
limiting the validity of our theory: (i) System sizes $L$ must be
{\em thermodynamically large} ($L\gg 2\pi\xi$),
so that Fourier sums over
${\bf k}$-space can be replaced by ${\bf k}$-integrals.
(ii) Times must be restricted to the {\em linear}
hydrodynamic regime ($\tau\lesssim\tau_{\rm cr}$), so that the system
remains close to the HCS.
Monitoring the energy per particle $E(t)$ provides a sensitive
criterion to distinguish the linear from
the nonlinear cooling regime, where the appearance of gradients
causes viscous heating and slows down the cooling \cite{goldhirsch} (see
Fig.\ 1b).
The crossover time $\tau_{\rm cr}$ in Fig.\ 1b, decreasing with
increasing $\epsilon$ and $\phi$, is an {\em intrinsic} time scale,
that only depends on the existing gradients.
At large $(\epsilon,\phi)$-values criterion (ii) reduces
$\tau_{\rm cr}(\epsilon;\phi)$
to subhydrodynamic time scales, with $\tau_{\rm cr}(0.5;0.25)\simeq
15$
and $\tau_{\rm cr}(0.3;0.4)\simeq 25$ as borderline cases.
On the other hand, small $(\epsilon,\phi)$-values combined with
small
$L$ tend to violate criterion (i).
The small systems with $N=5\times 10^3$, and to some extent even
those
with $N=2\times 10^4$, only satisfy the criteria (i) and (ii) in
very
narrow parameter ranges.
The systems studied in Ref.\ \cite{mcnamara} ($N=1024$),
\cite{deltour} ($N=1600$) and \cite{boston} ($N=5\times 10^3$)
are, in large regions of parameter space, so small that $L$ is
comparable to $2\pi\xi$, and the {\em periodic} boundaries
induce spurious transitions in the granular flows.
 
We have measured in simulations the equal time spatial correlation 
functions
$G_\mu(\br,t)$ with $\mu=\{nn,\parallel,\perp\}$ by two
methods,
first by summing $a_\mu({\bf v}_i) a_\mu({\bf v}_j)$ over
pairs of particles, where $a_\mu({\bf v}_i)=\{1,
({\bf v}_i\cdot\hat{\bf r}),({\bf v}_i\cdot\hat{\bf r}_\perp)
\}$, 
and binning their relative position vectors into
circular shells, and secondly by squaring the Fourier transform of
the coarse-grained fields,
followed by an inverse Fourier transformation.
The second approach also provides the correlations in ${\bf k}
$-space, which allows us to separate the contributions of
$S_\parallel(k,t)$ and $S_\perp(k,t)$ to $G_\parallel(r,t)$ and to
$G_\perp(r,t)$ and test the validity of the incompressibility
assumption.

\begin{figure}[h]
\[  \psfig{file=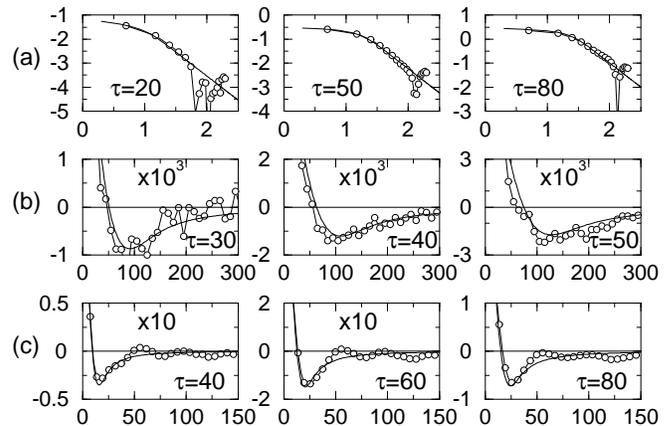,width=8.6cm} \]
\caption{
Correlation functions $G_\mu$ ($\mu=\{\parallel,\perp\}$) versus
$r$ at various times, inelasticities and densities
for $N=5\times 10^4$ particles, where simulation results of a
single run are compared with theoretical predictions (solid line).
Top row (a) shows $^{10}\log[G_\parallel/T]$ versus $^{10}\log{r}$
at $\alpha=0.9$ and $\phi=0.245$ ($l_0\simeq 0.8$),
where $\tau_{\rm cr}\simeq 70$.
The algebraic $1/r^2$-tail is clearly visible.
Middle row (b) shows $G_\perp/T$ versus $r$ at
$\alpha=0.94$ and $\phi=0.05$ ($l_0\simeq 5.8$)
where $\tau_{\rm cr}\simeq 100$.
Bottom row (c) shows $G_\perp/T$ versus $r$ at
$\alpha=0.9$ and $\phi=0.4$ ($l_0\simeq 0.34$)
where $\tau_{\rm cr}\simeq 65$.
Note regular oscillations with period $R_0\simeq 50$, which is
fixed in time.}
\end{figure}
 
Figures 2a,b,c show the longitudinal and transverse correlation
functions of the flow field of a {\em single} simulation run at
$N=5\times 10^4$ at several $(\epsilon,\phi)$-values.
The low noise levels, observed in these data and in Fig.\
 1a, are a consequence of the IHS collisions which have the
tendency to make particles move parallel.
There is reasonable agreement with the theoretical predictions from
our Langevin theory, in which the viscosity is taken from
Enskog's theory.
No fitting parameters are involved.
The longitudinal $G_\parallel(r,t)$ in Fig.\ 2a shows good
agreement for a large range of $(\epsilon,\phi)$-values, 
well beyond the linear
time regime $\tau_{\rm cr}$.  It exhibits the $1/r^2$-tail.
The minimum in $G_\perp(r,t)$ at $L_v(t)$ can be identified with
the mean vortex diameter, and the low noise data in Fig.\ 2c at
different $\tau$ show that $L_v(t)\sim \sqrt{\tau}$ is growing
through vorticity diffusion.
At small $(\epsilon,\phi)$-values $G_\perp(r,t)$
agrees well with our theory, as illustrated in Fig.\ 2b.
The $1/r^2$-tail in $G_\perp$ cannot be observed in a single run
because of
statistical fluctuations.
At larger densities (see Fig.\ 2c) one observes small oscillations
around the
predicted curve with a characteristic length $R_0\simeq 50$.
The oscillations become more pronounced at later times, where $R_0$
stays fixed in time, but varies over different runs.
Comparison of $G_\perp$ at $\tau=80$ with the snapshot in Fig.\ 1a
at the same parameters, suggests that
$G_\perp$ may be viewed as the pair correlation function of a
densely packed fluid of `hard objects' (vortices) of typical
diameter $L_v$, the oscillation length $R_0\simeq 2 L_v$ being
approximately equal to the size of a nearest neighbor $(+-)$-vortex
pair.
Similar complex structures, persisting in the nonlinear regime, are
typically observed at larger $(\epsilon,\phi)$-values ($N=4\times
10^4,
\epsilon=0.64, \phi=0.05$ \cite{goldhirsch}; $N=5\times 10^4,
\epsilon
\gtrsim 0.1, \phi\gtrsim 0.05$; $N=2\times 10^4, \epsilon >0.05,
\phi > 0.25$).
At smaller $(\epsilon,\phi)$-values (linear regime) the vortex
diameter
$L_v(t)\sim \sqrt{\tau}$, and a transition to a `sheared state' is
induced by the periodic boundaries when $L_v(t)\simeq
\textstyle{\frac{1}{2}}L$ \cite{boston}; for instance at
$\tau\simeq
600$ for $N=2\times 10^4, \epsilon=0.05, \phi=0.245$.

\begin{figure}
\[ \psfig{file=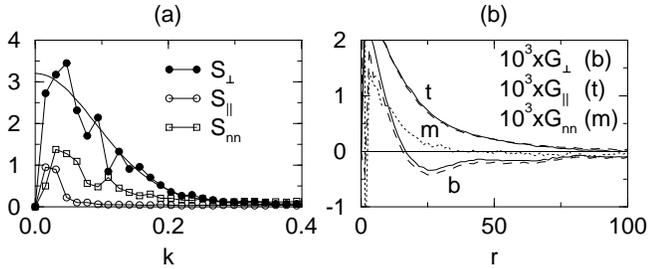,width=8.5cm} \]
\caption{
(a) Structure factors $S_\mu(k,t)$ ($\mu=\{\perp,\parallel,nn\}$) versus
$k$ from a single simulation run at $N=5\times 10^4$,
$\alpha=0.9$, $\phi=0.245$ and $\tau=50$ where $\tau_{\rm cr}\simeq 70$,
compared with the theoretical prediction (solid line) for $S_\perp
(k,t)$.
All structure in $S_\mu(k,t)$ contained in the interval
$k_{\rm min}=2\pi/L \simeq 0.016< k\ \protect{\lesssim}\  1/\xi=0.25$
represents long range correlations of dynamic origin.
(b) Correlation functions $G_\mu(r,t)\ (\mu=\{\parallel,\perp\})$
from $S_{\perp}$ and $S_\parallel$ (solid line), and from $S_\perp$ only
(dashed line).
$G_{nn}(r,t)$
(dotted line) corresponds to $S_{nn}(k,t)$.}
\end{figure}

The description of the velocity fluctuations
$G_{\alpha\beta}(\br,t)$ in this letter is based on fluctuating
hydrodynamics for the vorticity fluctuations only, i.e.\ the absence
of longitudinal fluctuations (incompressibility assumption).
Fig.\ 3a confirms that this assumption is very reasonable indeed,
as $S_\parallel(k,t)$ is vanishingly small down to very small
$k$-values ($k\gtrsim 1/\xi_\parallel \simeq
0.06$).
However for the smallest wavenumbers, the
incompressibility assumption breaks down.
In that range the longitudinal velocity fluctuations couple to the
second unstable mode \cite{mcnamara,deltour,boston} with a dispersion 
relation (to second order in
$k$)
$z_\parallel(k)=\gamma_0(1-k^2 \xi^2_\parallel)$.
The nonvanishing contributions of $S_\parallel(k,t)$ to
$G_{\alpha\beta}(\br,t)$ cause
an exponential cut-off on length scales $r\gtrsim  2\pi \xi_\parallel \gg 2
\pi \xi$, and
the algebraic decay $\sim 1/r^d$ from the vorticity mode represents
intermediate behavior, which is well observable because the two
length scales $\xi_\parallel$ and $\xi$ are in general quite
different, e.g.\ $\xi_\parallel \simeq 4.3\xi$
in Figs.\ 2a,c and $\xi_\parallel \simeq 4.5\xi$ in Fig.\ 2b,
and rapidly separate in the elastic limit.

The results of Fig.\ 3a were obtained by fast Fourier transformation of the
density and momentum fields (coarse-grained into $256\times 256$
cells), and performing an angular average in ${\bf k}$-space.
In the same figure one observes that $S_\parallel(k,t)$ has the
smallest width, while $S_\perp(k,t)$ and $S_{nn}(k,t)$ have a
comparable width.
Moreover, the growth rate of
$S_\perp(k,t)/E(t)$ exceeds that of $S_\parallel(k,t)/E(t)$, which
is in turn more unstable than $S_{nn}(k,t)$.
Finally, if one performs an inverse Fourier transform on the
{\em measured} $S_\perp(k,t)$ and $S_\parallel(k,t)$ separately to
obtain
the contributions to
$G_\parallel(r,t)$ and $G_\perp(r,t)$ (see Fig.\ 3b),
it appears that the contributions from $S_\parallel(k,t)$ are
small,
and our description of the fluctuations in terms of a Langevin equation 
based on incompressibility is confirmed by the 
simulations in the linear regime $\tau< \tau_{\rm cr}$.
\\

The authors want to thank H.J.~Bussemaker, D.~Frenkel, M.~Hagen and
W.~v.d.~Water for helpful comments and discussions. 
T.v.N. acknowledges support of the
foundation `Fundamenteel Onderzoek der Materie (FOM)', which is
financially supported by the Dutch National Science Foundation (NWO).
R.B. acknowledges support from DGICYT (Spain) number PB94-0265.

\end{multicols}

\begin{thebibliography}{99}
\bibitem{footnote} The same nonlinear
analysis of \cite{goldhirsch} reveals that the length scale
$\xi\sim l_0/\sqrt{\eps}$, associated with clustering, is the same
as the correlation length $\xi$ introduced above Eq.\ (4).
\end{thebibliography}
\end{document}